\RequirePackage{ifpdf}
\ifpdf 
\documentclass[pdftex]{sigma}
\else
\documentclass{sigma}
\fi

\begin{document}
\allowdisplaybreaks

\renewcommand{\PaperNumber}{062}

\FirstPageHeading

\ShortArticleName{Quantum Potential and Symmetries in Extended
Phase Space}

\ArticleName{Quantum Potential and Symmetries\\ in Extended Phase
Space}

\Author{Sadollah NASIRI~$^{\dag\ddag}$} 
\AuthorNameForHeading{S. Nasiri}

\Address{$^\dag$~Department of Physics, Zanjan University, Zanjan,
Iran} \Address{$^\ddag$~Institute for Advanced Studies in Basic
Sciences, IASBS, Zanjan, Iran}
\EmailD{\href{mailto:nasiri@iasbs.ac.ir}{nasiri@iasbs.ac.ir}}

\ArticleDates{Received January 06, 2006, in f\/inal form May 19,
2006; Published online June 27, 2006}

\Abstract{The behavior of the quantum potential is studied for a
particle in a linear and a~harmonic potential by means of  an
extended phase space technique. This is done by obtaining an
expression for the quantum potential in momentum space
representation followed by the generalization of this concept to
extended phase space. It is shown that there exists an extended
canonical transformation that removes the expression for the
quantum potential in the dynamical equation. The situation,
mathematically, is similar to disappearance of the centrifugal
potential in going from the spherical to the Cartesian coordinates
that changes the physical potential to an ef\/fective one. The
representation where the quantum potential disappears and the
modif\/ied Hamilton--Jacobi equation reduces to the familiar
classical form, is one in which the dynamical equation turns out
to be the Wigner equation.}


\Classification{81S30}

\section{Introduction}

If we assume a polar form for the wave function as the solution of
Schr\"odinger equation, we obtain decomposition of this equation
into real and imaginary parts \cite{holland1,sakurai}. The
resulting equations are a pair of coupled partial dif\/ferential
equations in which the amplitude and the phase of the wave
function co-determine each other. The real part, is nothing but
the familiar Hamilton--Jacobi equation with a term added that is
called the quantum potential, and the imaginary part
 yields the continuity equation for the amplitude \cite{madelung}. The
quantum potential term in the modif\/ied Hamilton--Jacobi equation
depends on the magnitude of the wave function rather than its
phase. This modif\/ication may be equally looked at from the point
of view of Newton second law that is modif\/ied by the quantum
force term. Thus, in the causal interpretation, in addition to the
external force, the quantum force, guides the trajectory of the
quantum particle in the framework of Bohm's approach to quantum
mechanics. Much has been done to identify the quantum aspects of
a~system within the quantum potential. The concept of quantum
internal energy (or stress) as a~consequence of the projection
from the phase space representation to the conf\/iguration space
representation was introduced by Takabayasi~\cite{takabayasi} for
the pure states and then extended by Muga et al.~\cite{muga} to the
mixed states. Unlike the classical systems that have kinetic and
potential energies, quantum systems also have intrinsic internal
energies associated with spatial localization and momentum
dispersion emerging from their inherent extended nature. It seems
that the quantum potential is a balance between the energies of
spatial localization and momentum dispersion in quantum systems
suggesting a link to the Heisenberg position-momentum uncertainty
principle~\cite{brown1}. Brown and Hiley~\cite{brown2} showed that
the quantum potential cannot be dismissed as an artif\/icial term
without missing some of the essential novel features of quantum
mechanics accounting for the interference,
Einstein--Podolsky--Rosen type correlations~\cite{bohm1}, quantum
state teleportation~\cite{maroney}, barrier penetration and
quantum non-separability~\cite{bohm2}. Holland~\cite{holland2},
has recently investigated the de Broglie--Bohm law of motion using
a variational formulation. He considers a quantum system and
suggests a total Lagrangian for the interaction of a point like
particle with Schr\"odinger f\/ield without a back reaction of the
particle on the wave. The interaction of the particle and the
f\/ield is attributed to a scalar potential that turns out to be
quantum potential, whenever, the constraint of the invariance of
Schr\"odinger equation is applied. The connection of quantum
potential with quantum f\/luctuations and quantum geometry in
terms of Weyl curvature has been studied extensively by
dif\/ferent authors; see, for example F.~and~A.~Shojai
\cite{shijai}, Carroll~\cite{carroll1} and the references therein.
However, unlike the external potential, the quantum potential is
not a pre-assigned function of the system coordinates and can only
be derived from the wave function of the system~\cite{brown1} or
from the corresponding quantum distribution functions used to
calculate the average values of the observables. This
representation dependent property of the quantum potential, as the
main theme of the present work, allows one to f\/ind appropriate
representations in which one may obtain desired expressions for
the quantum potential that may even be equal to zero. In an
interesting work by Carroll~\cite{carroll2} it is shown that for a
quite general function of position, time and $\hbar$,
$Q(q,t,\hbar)$, there are ``generalized'' quantum theories having
$Q$ as their quantum potential and that the vanishing of the
quantum potential depends on the wave function.

Here, we show that in an appropriate phase space representation
one may remove the quantum potential from the dynamical equation
of a particle in the linear and the harmonic potential keeping the
Hamilton--Jacobi equation form invariant. In other words, we show
that the quantum potential behaves as a representation dependent
quantity in the phase space. The situation seems to be similar to
the case of centrifugal potential that disappears in going from
spherical to Cartesian coordinate system. The representation
 is obtained by a canonical transformation on the classical level corresponding to a
unitary transformation on the quantum level.

In 1932 the phase space picture of the quantum mechanics was
introduced by the pioneering work of Wigner~\cite{wigner} and
extended by Moyal~\cite{moyal}, Hillary et al.~\cite{hillary},
Mehta \cite{mehta}, Agarwal and Wolf \cite{agarval}, Han et
al.~\cite{han} Kim and Wigner \cite{kim} and  Jannussis et
al.~\cite{jannussis}, and many others. Although, the arguments
invoked on the basis of the Heisenberg uncertainty principle made
the physical meaning of the "phase space points" problematic,
things have changed and phase space techniques mainly formulated
by the theory of deformation quantization \cite{bayen} and a
family of Schr\"odinger equations in phase space
\cite{tores1,tores2,gosson1,bolivar,gosson2} are now widely
accepted and used. As an approach of the latter type, Sobouti and
Nasiri \cite{sobouti&nasiri} have proposed a formulation of
quantum statistical mechanics by generalizing the principle of
least action to the trajectories in phase space, and a canonical
quantization procedure in this space. In the extended phase space
(EPS) formalism, positions and momenta are assumed to be
independent c-number variables. Therefore, the canonical
transformations allow one to def\/ine dif\/ferent phase space
representations for quantum distribution functions with their
corresponding associated ordering rules. As a~special case, we
obtain the Wigner distribution functions \cite{moyal,mehta} by a
simple extended canonical transformation that corresponds to an
extended unitary transformation at quantum level.

Using the EPS method, we look for a canonical transformation that
removes the quantum potential in the dynamical equation. In
Section~2, the EPS formalism that treats the momentum and position
coordinates in a symmetric way after quantization, is reviewed. In
Section~3, the extended canonical transformations are introduced
and are used to obtain the Wigner equation and the corresponding
Wigner functions. In Section~4, the quantum potential is written
in both conf\/iguration and momentum space representations for
linear and harmonic potentials. It is argued that the form of the
quantum potential in the momentum space is not, in general, as
simple as that of the conf\/iguration space. Furthermore, an
expression for the quantum potential is obtained in EPS. In
Section~5, an appropriate phase space representation, which is
shown to be the Wigner representation of quantum mechanics in the
phase space \cite{wigner}, is found in which the quantum potential
is removed in dynamical equations for linear and harmonic
potentials. Finally, Section~6 is devoted to the concluding
remarks.

\section{Review of EPS formalism}

A direct approach to quantum statistical mechanics is proposed by
extending the conventional phase space and by applying the
canonical quantization procedure to the extended quantities in
this space. Here, a brief review of this formalism is presented.
For more details, see Sobouti and Nasiri \cite{sobouti&nasiri}.

Let ${\cal L}^q(q,\dot{q})$ be a Lagrangian specifying a system in
$q$ space. A trajectory of the system in this space is obtained by
solving the Euler--Lagrange equations for $q(t)$,
\begin{gather}\label{eq1}
\frac{d}{dt}\frac{\partial {\cal L}^q}{\partial \dot
q}-\frac{\partial {\cal L}^q}{\partial q}=0.
\end{gather}
The derivative $\frac{\partial {\cal L}^q}{\partial \dot q}$
calculated on an actual trajectory, that is, on a~solution of
equation~\eqref{eq1}, is the momentum $p$ conjugate to $q$. The
same derivative calculated on a virtual orbit, not a~solution of
equation~\eqref{eq1}, exists. It may not, however, be interpreted
as a canonical momentum. Let $H(p,q)$ be a function in phase space
which is the Hamiltonian of the system, whenever $p$ and $q$ are
canonical pairs. It is related to ${\cal L}^q$ through the
Legendre transformation,
\begin{gather}\label{eq2}
{H}\left(\frac{\partial {\cal L}^q}{\partial \dot
q},q\right)=\dot{q}\frac{\partial {\cal L}^q}{\partial \dot
q}-{\cal L}^q(q,\dot q).
\end{gather}
For a given ${\cal L}^q$, equation~\eqref{eq2} is an algebraic
equation for $H$. One may, however, take a dif\/ferent point of
view. For a given functional form of $H(p,q)$,
equation~\eqref{eq2} may be considered as a~dif\/fe\-ren\-tial
equation for ${\cal L}^q$. Its non unique solutions dif\/fer from
one another by total time derivatives. One may also study the same
system in the momentum space. Let ${\cal L}^p(p,\dot p)$ be
a~Lagrangian in $p$ space \cite{goldstein}. It is related to
$H(p,q)$ as follows,
\begin{gather}\label{eq3}
H\left(p,\frac{\partial {\cal L}^p}{\partial \dot
p}\right)=-\dot{p}\frac{\partial {\cal L}^p}{\partial \dot
p}+{\cal L}^p(p,\dot p).
\end{gather}
Here the functional dependence of $H$ on its argument is the same
as in equation~\eqref{eq2}. In principle, equation~\eqref{eq3}
should be solvable for ${\cal L}^p$ up to an additive total time
derivative term. Once ${\cal L}^p$ is known the actual
trajectories in $p$ space are obtainable from an Euler--Lagrange
equation analogous to equation~\eqref{eq1} in which  $q$ is
replaced by $p$. That is
\begin{gather*}
\frac{d}{dt}\frac{\partial {\cal L}^p}{\partial \dot
p}-\frac{\partial {\cal L}^p}{\partial p}=0.
\end{gather*}
The derivative $\frac{\partial {\cal L}^p}{\partial \dot p}$ along
an actual $p$ trajectory is the canonical coordinate conjugate to
$p$. Calculated on a virtual orbit, it is not.

A formulation of quantum statistical mechanics is possible by
extending the conventional phase space and by applying the
canonical quantization procedure to the extended quantities in
this space. Assuming the phase space coordinates $p$ and $q$ to be
independent variables on the virtual trajectories, allows one to
def\/ine momenta $\pi_{p}$ and $\pi_{q}$, conjugate to $p$ and
$q$, respectively. One may combine the two pictures and def\/ine
an extended Lagrangian in the phase space as the sum of $p$ and
$q$ Lagrangians,
\begin{gather}\label{eq5}
{\cal L}(p,q,{\dot p},{\dot q})=-{\dot p}q-{\dot q}p + {\cal L}^{p}(p,{\dot p})+ {\cal L}^{q}(q,{\dot q}).
\end{gather}
The f\/irst two terms in equation~\eqref{eq5} constitute a total
time derivative. The equations of motion are
\begin{gather}\frac{d}{dt}\frac{\partial {\cal L}}{\partial \dot p}-\frac{\partial
{\cal L}}{\partial p}=\frac{d}{dt}\frac{\partial {\cal
L}^p}{\partial \dot p}-\frac{\partial {\cal L}^p}{\partial
p}=0,\nonumber
\\
\frac{d}{dt}\frac{\partial {\cal L}}{\partial \dot
q}-\frac{\partial {\cal L}}{\partial q}=\frac{d}{dt}\frac{\partial
{\cal L}^q}{\partial \dot q}-\frac{\partial {\cal L}^q}{\partial
q}=0.\label{eq6}
\end{gather}

The $p$ and $q$ in equations~\eqref{eq5} and \eqref{eq6} are not,
in general, canonical pairs. They are so only on actual
trajectories and through a proper choice of the initial values.
This gives the freedom of introducing a second set of canonical
momenta for both $p$ and $q$. One does this through the extended
Lagrangian. Thus
\begin{gather*}
\pi_{p} = \frac{\partial {\cal  L} }{\partial {\dot p} } =
\frac{\partial {\cal L}^p}{\partial {\dot p}}-q, \qquad \pi_{q} =
\frac{\partial {\cal  L }}{\partial {\dot q} } = \frac{\partial
{\cal L} ^q}{\partial {\dot q} }-p.
\end{gather*}
Evidently, $\pi_p$ and $\pi_q$ vanish on actual trajectories and
remain non zero on virtual ones. From these extended momenta, one
def\/ines an extended Hamiltonian,
\begin{gather}
{\cal H}(\pi_p ,\pi_q , p,q)  =  {\dot p}\pi_p +{\dot q}\pi_q -{\cal L}= H(p+\pi_q,q)-H(p,q+\pi_p)\nonumber\\
\phantom{{\cal H}(\pi_p ,\pi_q , p,q)}{} =
\sum\frac{1}{n!}\left\{\left(\frac{\partial^nH}{\partial
p^n}\right)_{|_{\pi_q=0}}\pi_{q}^n
-\left(\frac{\partial^nH}{\partial
q^n}\right)_{|_{\pi_p=0}}\pi_{p}^n\right\}.\label{eq8}
\end{gather}

By means of utilisation of the canonical quantization rule, the
following postulates are outlined:

a) Let $p$, $q$, $\pi_p$ and $\pi_q$ be operators in a Hilbert
space, $\bf X$, of all complex functions, satisfying the following
commutation relations
\begin{gather}
[\pi_{q},q]=-i\hbar,\qquad \pi_{q}=-i\hbar\frac{\partial}{\partial
q},
\nonumber\\
[\pi_{p},p]=-i\hbar,\qquad \pi_{p}=-i\hbar\frac{\partial}{\partial
p}, \qquad [p,q]=[\pi_{p},\pi_{q}]=0.\label{eq9}
\end{gather}
By virtue of equations~\eqref{eq9}, the extended Hamiltonian,
${\cal H}$, will also be an operator in ${\bf X}$.

b) A state function $\chi(p,q,t)\in{\bf X}$ is assumed to satisfy
the following dynamical equation
\begin{gather}
i\hbar\frac{\partial \chi}{\partial t}={\cal
H}\chi=\left[H\left(p-i\hbar\frac{\partial}{\partial q},q\right)
-H\left(p,q-i\hbar\frac{\partial}{\partial p}\right)\right]\chi\nonumber\\
\phantom{i\hbar\frac{\partial \chi}{\partial t}={\cal H}\chi}{}
=\sum\frac{(-i\hbar)^n}{n!}\left\{\frac{\partial^nH}{\partial
p^n}\frac{\partial^n}{\partial q^n}-\frac{\partial^nH}{\partial
q^n} \frac{\partial^n}{\partial p^n}\right\}\chi.\label{eq10}
\end{gather}

The evolution operator on the right hand side of
equation~\eqref{eq10} is Hermitian as far as the Hamiltonian $H$
is Hermitian and the extension operation is done by the canonical
transformations corresponding to the unitary transformations at
quantum level. Otherwise, it is not Hermitian and the evolution as
described is not unitary. In this case one may wonder that the
probability conservation may be violated. However, it is known
that $\chi$, unlike the classical Liouville equation, is not a
probability distribution function. It may become negative or even
complex in some regions of the phase space.

c) The averaging rule for an observable $O(p,q)$, a $c$-number
operator in this formalism, is given as
\begin{gather*}
\langle O(p,q)\rangle=\int O(p,q)\chi^{*}(p,q,t)\, dp\, dq.
\end{gather*}
To f\/ind the solutions for equation~\eqref{eq10} one may assume
\begin{gather}\label{eq12}
\chi(p,q,t)=F(p,q,t)e^{-ipq/\hbar}.
\end{gather}

The phase factor comes out due to the total derivative in the
Lagrangian of equation~\eqref{eq5}, $-d(pq)/dt$. The ef\/fect is
the appearance of a phase factor, $\exp(-ipq/\hbar )$, in the
state function that would have been in the absence of the total
derivative. It is easily verif\/ied that
\begin{gather}
\left(p-i\hbar\frac{\partial}{\partial
q}\right)\chi=i\hbar\frac{\partial F}{\partial q}e^{-ipq/\hbar},
\qquad \left(q-i\hbar\frac{\partial}{\partial
p}\right)\chi=i\hbar\frac{\partial F}{\partial
p}e^{-ipq/\hbar}.\label{eq13}
\end{gather}

Substituting of equations \eqref{eq13} into equation~\eqref{eq10}
and eliminating of the exponential factor gives
\begin{gather}\label{eq14}
{H\left(-i\hbar\frac{\partial}{\partial q},q\right)
-H\left(p,-i\hbar\frac{\partial}{\partial
p}\right)}F=i\hbar\frac{\partial F}{\partial t}.
\end{gather}
Equation \eqref{eq14} has separable solutions of the form
\begin{gather}\label{eq15}
F(p,q,t)=\psi(q,t)\phi^{\ast}(p,t),
\end{gather}
where $\psi(q,t)$ and $\phi(p,t)$ are the solutions of the
Schr\"odinger equation in $q$ and $p$ representations,
respectively. The solution of the form \eqref{eq12} associated
with anti-standard ordering rule satisf\/ies equation~\eqref{eq10}
and is one possible distribution function. For more details on the
admissibility of the above distribution function, their
interesting properties and the correspondence rules, the
interested reader may consult Hillary et al.~\cite{hillary} and
Sobouti and Nasiri \cite{sobouti&nasiri}.

\section{The extended canonical transformations}
The canonical transformations that keeps the extended Hamilton
equation of motion form invariant are possible in EPS
\cite{sobouti&nasiri}. Consider an inf\/initesimal extended
canonical transformation on $p$, $q$, $\pi_p$, $\pi_q$ as follows
\begin{gather}\label{eq16}
p  \rightarrow  p+\alpha\pi_q,\qquad
 q  \rightarrow q+\beta \pi_p,\qquad \pi_p  \rightarrow  \pi_p+\gamma q,\qquad
 \pi_q  \rightarrow  \pi_q+\eta p.
\end{gather}

The above transformation is prepared in such a way that the
ordering problem does not exist according to the commutation
relations of equation~\eqref{eq9}. The transformation to be
canonical, one must have \cite{goldstein}
\begin{gather*}
 -\beta+\alpha=0,\qquad
 -\beta\eta+1=1,\qquad
-1+\alpha\gamma=-1.
\end{gather*}
The result is $\beta=\alpha$ and $\eta=\gamma=0$. The
corresponding generator is \cite{merzbacher}
\begin{gather*}
G=\pi_p\pi_q= -\hbar^2\frac{\partial^2}{\partial p\partial q}.
\end{gather*}
Then, the unitary transformation for f\/inite $\alpha$ becomes
\begin{gather}\label{eq19}
U=e^{\alpha G/i\hbar}=e^{-i\alpha\hbar\partial^2/{\partial
p\partial q}}=e^{-i\alpha\hbar\partial^2/{\partial P\partial Q}},
\qquad UU^{\dagger}=1.
\end{gather}

By means of utilisation of equations~\eqref{eq8} and \eqref{eq19}
for $\alpha=-1/2$, the new extended Hamiltonian becomes
\begin{gather}\label{eq20}
{\cal H}'= U{\cal H}U^{-1}=
H\left(p+\frac{1}{2}\pi_q,q-\frac{1}{2}\pi_p\right)-H\left(p-\frac{1}{2}\pi_q,q+\frac{1}{2}\pi_p\right).
\end{gather}
For $H=\frac {1}{2m} p^2+V(q)$, equation~\eqref{eq20} gives
\begin{gather}\label{eq21}
{\cal H}' = \frac
{1}{2m}{\left(p+\frac{1}{2}\pi_q\right)}^2+V\left(q-\frac{1}{2}\pi_p\right)-\frac
{1}{2m}{\left(p-\frac{1}{2}\pi_q\right)}^2-V\left(q+\frac{1}{2}\pi_p\right).
\end{gather}
As explicit examples we examine the above transformation for a
particle in the linear and harmonic potentials. For the linear
potential, $H=\frac {1}{2m} p^2+bq$, $b$ a constant, one gets
simply
\begin{gather}\label{eq22}
{\cal H}' = \frac {p}{m}\pi_q-b\pi_p,
\end{gather}
while, for harmonic potential, $H=\frac {1}{2m}
p^2+\frac{1}{2}kq^2$, $k=m\omega^2$, one gets simply
\begin{gather}\label{eq23}
{\cal H}' = \frac {p}{m}\pi_q-kq\pi_p.
\end{gather}

By means of utilisation of equation \eqref{eq21}, the evolution
equation \eqref{eq10} and the distribution functions $\chi(p,q,t)$
transform to
\begin{subequations}\label{eq24}
\begin{gather}\label{eq24a}
i\hbar\frac{\partial W}{\partial t}= {\cal
H}'W=-i\hbar\frac{p}{m}\frac{\partial W}{\partial q} + \sum
\frac{i\hbar}{(2n+1)!}\left({\frac{\hbar}{2i}}\right)^{2n}
\frac{\partial^{2n+1}V}{\partial
q^{2n+1}}\frac{\partial^{2n+1}W}{\partial p^{2n+1}}
\end{gather}
and
\begin{gather}\label{eq24b}
W(p,q,t)= U\chi= \int \psi\left(q+\frac{1}{2}\hbar\tau,t\right)
\psi^*\left(q-\frac{1}{2}\hbar\tau,t\right)e^{-ip\tau}d\tau,
\end{gather}
\end{subequations}
respectively. Equation \eqref{eq24a} is the Wigner equation
governing the Wigner distribution functions given by
equation~\eqref{eq24b} \cite{carroll1}. Therefore, we call the new
form of the EPS obtained by transformation \eqref{eq16} on the old
one for $\beta=\alpha=-\frac{1}{2}$ as the Wigner representation.

\section{Quantum potential and generalization to the EPS}

\textbf{a) Quantum potential in $\boldsymbol{q}$ space.} Let the
Schr\"odinger wave function in $q$ space be written as follows
\cite{holland1}
\begin{gather}\label{eq25}
\psi(q,t)=R^q(q,t)e^{iS^q(q,t)/\hbar},
\end{gather}
where $R^q(q,t)$ is the amplitude and $S^q(q,t)$ is the action
that serves as the phase of the wave function in $q$ space and is
\begin{gather}\label{eq26}
S^q(q,t)=\int^t {\cal L}^q(q, \dot q,t')dt'.
\end{gather}

Using equation~\eqref{eq25} in the Schr\"odinger equation one gets
\begin{gather}
i\hbar\left(\frac{\partial R^q}{\partial
t}+\frac{i}{\hbar}R^q\frac{\partial S^q}{\partial t}\right)\nonumber\\
\qquad{}= -\frac{\hbar^2}{2m}\left[\frac{\partial^2R^q}{\partial
q^2}+\frac{2i}{\hbar}\frac{\partial R^q}{\partial q}\frac{\partial
S^q}{\partial q}+ \frac{i}{\hbar}R^q\frac{\partial^2S^q}{\partial
q^2} - \frac{1}{\hbar^2}R^q\left(\frac{\partial S^q}{\partial
q}\right)^2\right]+V(q)R^q.\label{eq27}
\end{gather}
The real part of equation~\eqref{eq27} gives
\begin{gather}\label{eq28}
\frac{\partial S^q}{\partial
t}-\frac{\hbar^2}{2m}\frac{1}{R^q}\frac{\partial^2R^q}{\partial
q^2}+\frac{1}{2m}\left(\frac{\partial S^q}{\partial
q}\right)^2+V(q)=0.
\end{gather}
Apart from the term
$-\frac{\hbar^2}{2m}\frac{1}{R^q}\frac{\partial^2R^q}{\partial
q^2}$, equation~\eqref{eq28} for $p=\frac{\partial S^q}{\partial
q}$ is the familiar Hamilton--Jacobi equation. The term
$-\frac{\hbar^2}{2m}\frac{1}{R^q}\frac{\partial^2R^q}{\partial
q^2}$ is known as the quantum potential \cite{holland1}. The
imaginary part of equation~\eqref{eq27} yields a continuity
equation for $|R|^2$, that is not our interest here.

\textbf{b) Quantum potential in $\boldsymbol{p}$ space.} The
expression for quantum potential in $p$ space does not, in
general, have a simple one term form as in equation~\eqref{eq27}
obtained for $q$ space~\cite{brown1}. However, the cases of linear
and harmonic potentials are relatively simple. Potentials of
higher polynomial order generate series of terms having even
powers of $\hbar^2$. See~\cite{brown1} for some explicit examples.
Here, we concentrate on the cases of linear and harmonic
potentials.

i) \emph{Linear potential.} The Schr\"odinger equation for a
particle in a linear potential of $V(q)=bq$ in $p$ representation
is
\begin{gather}\label{eq29}
i\hbar \frac{\partial \phi(p,t)}{\partial
t}=\frac{p^2}{2m}\phi(p,t)+i\hbar b\frac{\partial
\phi(p,t)}{\partial p},
\end{gather}
 where $\phi(p,t)$ is the Schr\"odinger wave
function in $p$ space and is assumed to be
\begin{gather}\label{eq30}
\phi(p,t)=R^p(p,t)e^{-iS^p(p,t)/\hbar},
\end{gather}
where $R^p(p,t)$ is the amplitude and $S^p(p,t)$ is def\/ined as
\begin{gather}\label{eq31}
S^p(p,t)=\int^t {\cal L}^p(p, \dot p,t')dt'
\end{gather}
where ${\cal L}^p$ is given by equation~\eqref{eq3}.

Substituting of equation \eqref{eq30} into \eqref{eq29} gives
\begin{gather}\label{eq32}
i\hbar\left(\frac{\partial R^p}{\partial
t}+\frac{i}{\hbar}R^p\frac{\partial S^p}{\partial t}\right)=
\frac{p^2}{2m}R^p + i\hbar b\left(\frac{\partial R^p}{\partial
p}+\frac{iR^p}{\hbar}\frac{\partial S^p}{\partial p}\right).
\end{gather}
The real part of equation \eqref{eq32} gives
\begin{gather*}
\frac{\partial S^p}{\partial t}+\frac{p^2}{2m} - b\frac{\partial
S^p}{\partial p}=0.
\end{gather*}
Identifying $q=-\frac{\partial S^p}{\partial p}$, gives
\begin{gather*}
\frac{\partial S^p}{\partial t}+H=0.
\end{gather*}

Note that, unlike the $q$ space, there is no quantum potential
term for the linear potential in $p$ space \cite{brown1}. In other
words, the quantum potential is removed in going from the $q$
space to the $p$ space. Thus, if one works out the problem of a
particle in the linear potential in the $p$ space, quantum
potential will not appear in the dynamical equations.

ii) \emph{Harmonic potential.} By the same procedure followed in
part (i), one gets for harmonic potential $V(q)= \frac{1}{2}kq^2$,
\begin{gather}
i\hbar\left(\frac{\partial R^p}{\partial
t}+\frac{i}{\hbar}R^p\frac{\partial S^p}{\partial t}\right)\nonumber\\
\qquad{}=
\frac{p^2}{2m}R^p-\frac{\hbar^2}{2}k\left[\frac{\partial^2R^p}{\partial
p^2}+\frac{2i}{\hbar}\frac{\partial R^p}{\partial p}\frac{\partial
S^p}{\partial p}+ \frac{i}{\hbar}R^p\frac{\partial^2S^p}{\partial
p^2}
 - \frac{1}{\hbar^2}R^p\left(\frac{\partial S^p}{\partial
p}\right)^2\right].\label{eq35}
\end{gather}
The real part of equation~\eqref{eq35} gives
\begin{gather}\label{eq36}
\frac{\partial S^p}{\partial
t}+\frac{p^2}{2m}-\frac{\hbar^2}{2}\frac{k}{R^p}\frac{\partial^2R^p}{\partial
p^2}+\frac{1}{2}k\left(\frac{\partial S^p}{\partial p}\right)^2=0.
\end{gather}
Equation \eqref{eq36} is the modif\/ied Hamilton--Jacobi equation
for the harmonic potential in the $p$ space and contains an extra
term,
$-\frac{\hbar^2}{2}\frac{k}{R^p}\frac{\partial^2R^p}{\partial
p^2}$, as $p$ space version of the quantum potential.

\textbf{c) Quantum potential in the EPS.} To generalize the
concept of quantum potential into the EPS we again consider the
cases of the linear and harmonic potentials as we did in part {\bf
b}.

i) \emph{Linear potential.} The extended Hamiltonian of
equation~\eqref{eq8} for the linear potential becomes
\begin{gather*}
{\cal H}=\frac{\pi_q^2}{2m}+\frac{p}{m}\pi_q-b\pi_p.
\end{gather*}
The state function, $\chi$, in equation~\eqref{eq10} is, in
general, a complex function. Thus, we assume
\begin{gather}\label{eq38}
\chi(p,q,t)={\cal R}(p,q,t)e^{i{\cal S}(p,q,t)/\hbar},
\end{gather}
where ${\cal R}(p,q,t)$ is the amplitude and ${\cal S}(p,q,t)$ is
the phase given by
\begin{gather*}
{\cal S}(p,q,t)=\int^t {\cal L}(p, q, \dot p,\dot q,t')dt',
\end{gather*}
where ${\cal L}(p, q, \dot p,\dot q,t)$ is def\/ined by
equation~\eqref{eq5}.

Using equation~\eqref{eq5}, one gets
\begin{gather}\label{eq40}
{\cal S}(p,q,t)=S^p+S^q-pq,
\end{gather}
where $S^q$ and $S^p$ are given by equations~\eqref{eq26}
and~\eqref{eq31}, respectively. An alternative approach to
equation~\eqref{eq40} is to use equations~\eqref{eq25}
and~\eqref{eq30} in equations~\eqref{eq12} and \eqref{eq15} and
compare the result with equation~\eqref{eq38}. One may also see
that ${\cal R}(p,q,t)=R^q(q,t)R^{*p}(p,t)$.

Equation \eqref{eq10} now gives
\begin{gather}
i\hbar\left(\frac{\partial {\cal R}}{\partial t} +\frac{i{\cal
R}}{\hbar}\frac{\partial {\cal S}}{\partial t}\right)
=-\frac{\hbar^2}{2m}\left[\frac{\partial^2{\cal R}}{\partial
q^2}+\frac{2i}{\hbar}\frac{\partial {\cal R}}{\partial
q}\frac{\partial {\cal S}}{\partial q} +\frac{i{\cal
R}}{\hbar}\frac{\partial^2{\cal S}}
{\partial q^2}-\frac{{\cal R}}{\hbar^2}\left(\frac{\partial{\cal S}}{\partial q}\right)^2\right]\nonumber\\
\phantom{i\hbar\left(\frac{\partial {\cal R}}{\partial t}
+\frac{i{\cal R}}{\hbar}\frac{\partial {\cal S}}{\partial
t}\right)=}{}
 -  \frac{i\hbar p}{m}\left(\frac{\partial {\cal R}}{\partial
q}+\frac{i{\cal R}}{\hbar}\frac{\partial {\cal S}}{\partial
q}\right)+
 i\hbar b\left(\frac{\partial {\cal R}}{\partial
p}+\frac{i{\cal R}}{\hbar}\frac{\partial {\cal S}}{\partial
p}\right).\label{eq41}
\end{gather}
With assumption
\begin{gather*}
\pi_p=\frac{\partial {\cal S}(p,q,t)}{\partial p} \qquad
\mbox{and} \qquad \pi_q=\frac{\partial {\cal S}(p,q,t)}{\partial
q},
\end{gather*}
the real part  of equation~\eqref{eq41} gives
\begin{gather}\label{eq44}
\frac{\partial {\cal S}}{\partial
t}-\frac{\hbar^2}{2m}\frac{1}{{\cal R}}\frac{\partial^2{\cal
R}}{\partial q^2}+{\cal H}=0.
\end{gather}
Equation \eqref{eq44} is the modif\/ied Hamilton--Jacobi equation
for the linear potential in EPS.

ii) \emph{Harmonic potential.} The extended Hamiltonian for
harmonic potential is
\begin{gather*}
{\cal
H}=\frac{\pi_q^2}{2m}+\frac{p}{m}\pi_q-\frac{1}{2}k\pi^2_p-kq\pi_p.
\end{gather*}
The corresponding evolution equation becomes
\begin{gather}
i\hbar\left(\frac{\partial {\cal R}}{\partial t} +\frac{i{\cal
R}}{\hbar}\frac{\partial {\cal S}}{\partial t}\right)
=-\frac{\hbar^2}{2m}\left[\frac{\partial^2{\cal R}}{\partial
q^2}+\frac{2i}{\hbar}\frac{\partial {\cal R}}{\partial
q}\frac{\partial {\cal S}}{\partial q} +\frac{i{\cal
R}}{\hbar}\frac{\partial^2{\cal S}}
{\partial q^2}-\frac{{\cal R}}{\hbar^2}\left(\frac{\partial{\cal S}}{\partial q}\right)^2\right]\nonumber\\
\qquad{}-  \frac{i\hbar p}{m}\left(\frac{\partial {\cal
R}}{\partial q}+\frac{i{\cal R}}{\hbar}\frac{\partial {\cal
S}}{\partial q}\right) +  \frac{k\hbar^2}{2}\left[\frac{\partial^2
{\cal R}}{\partial p^2}+\frac{2i}{\hbar}\frac{\partial {\cal
R}}{\partial p}\frac{\partial {\cal S}}{\partial p}+\frac{i{\cal
R}}{\hbar}\frac{\partial^2{\cal S}}
{\partial p^2}-\frac{{\cal R}}{\hbar^2}\left(\frac{\partial {\cal S}}{\partial p}\right)^2\right]\nonumber\\
\qquad{}+  i\hbar k q\left(\frac{\partial {\cal R}}{\partial
p}+\frac{i{\cal R}}{\hbar}\frac{\partial {\cal S}}{\partial
p}\right).\label{eq46}
\end{gather}
The real part  of equation~\eqref{eq46} gives
\begin{gather}
\frac{\partial {\cal S}}{\partial
t}-\frac{\hbar^2}{2m}\frac{1}{{\cal R}}\frac{\partial^2{\cal
R}}{\partial q^2}+\frac{\hbar^2k}{2}\frac{1}{{\cal
R}}\frac{\partial^2{\cal R}}{\partial p^2}+{\cal H}=0.\label{eq47}
\end{gather}
Equation \eqref{eq47} is the modif\/ied Hamilton--Jacobi equation
for harmonic potential in EPS. The second and the third terms in
equation~\eqref{eq47}, together, def\/ine the quantum potential in
extended phase space. The second term is the EPS counterpart of
quantum potential in $q$ space and the third term is the same
thing in $p$ space.

\section{How to remove the quantum potential}
We look for a possible extended canonical transformation that
could remove the quantum potential from the modif\/ied
Hamilton--Jacobi equations \eqref{eq44} and \eqref{eq47}. In this
respect we consider the following linear transformation
\begin{gather}\label{eq48}
p  \rightarrow  p+\alpha\pi_q,\qquad  q  \rightarrow q+\alpha
\pi_p,\qquad \pi_p  \rightarrow  \pi_p,\qquad  \pi_q  \rightarrow
\pi_q.
\end{gather}

As argued in Section 3, transformation \eqref{eq48} is canonical
for any arbitrary $\alpha$. Thus, the new extended Hamiltonians
for the linear and the harmonic potentials become
\begin{gather*}
{\cal
H}'=\frac{\pi_q^2}{2m}+\frac{(p+\alpha\pi_q)}{m}\pi_q-b\pi_p,
\end{gather*}
and
\begin{gather*}
{\cal
H}'=\frac{\pi_q^2}{2m}+\frac{(p+\alpha\pi_q)}{m}\pi_q-\frac{1}{2}k\pi^2_p-
k(q+\alpha\pi_p)\pi_p,
\end{gather*}
respectively.

It can be easily shown that the modif\/ied Hamilton--Jacobi
equations \eqref{eq44} and \eqref{eq47} transform to
\begin{gather}\label{eq51}
\frac{\partial {\cal S'}}{\partial
t}-\frac{\hbar^2}{m}\left(\frac{1}{2}+\alpha\right)\frac{1}{{\cal
R'}}\frac{\partial^2{\cal R'}}{\partial q^2}+\frac
{p}{m}\pi_q-b\pi_p=0,
\end{gather}
and
\begin{gather}\label{eq52}
\frac{\partial {\cal S'}}{\partial
t}-\frac{\hbar^2}{m}\left(\frac{1}{2}+\alpha\right)\frac{1}{{\cal
R'}}\frac{\partial^2{\cal R'}}{\partial
q^2}+\hbar^2k\left(\frac{1}{2}+\alpha\right)\frac{1}{{\cal
R'}}\frac{\partial^2{\cal R'}}{\partial p^2}+\frac
{p}{m}\pi_q-kq\pi_p=0,
\end{gather}
respectively.

With assumption $\alpha=-\frac{1}{2}$ and by means of utilisation
of equations \eqref{eq22} and \eqref{eq23}, both of
equations~\eqref{eq51} and \eqref{eq52} simultaneously become
\begin{gather}\label{eq53}
\frac{\partial {\cal S'}}{\partial t}+{\cal H'}=0.
\end{gather}
Equation \eqref{eq53} has the familiar form of classical
Hamilton--Jacobi equation. Note that, as claimed before,
equation~\eqref{eq16} transforms equation~\eqref{eq10}
 into Wigner equation for $\alpha=-\frac{1}{2}$ and
$\eta=\gamma=0$. Thus, we conclude that equation~\eqref{eq53} in
which the quantum potential is disappeared is obtained in the
Wigner representation. The generalization of the present technique
to the potentials having a multinomial expansion is under
consideration. So far, it is already shown that one may take away
the quantum potential from the modif\/ied Hamilton--Jacobi
equation to the leading terms proportional to the higher powers of
$\hbar$ and remove the quantum potential up to the desired
approximation order.

\section{Conclusions}

In this paper the concept of quantum potential in conf\/iguration
space is generalized to the momentum space as well as the phase
space. The modif\/ied Hamilton--Jacobi equation containing the
quantum potential is obtained in these spaces by substituting a
polar form of the distribution function into the corresponding
dynamical equation. By means of utilisation of the EPS formalism,
it is shown that for the cases of linear and harmonic potentials
one may remove the quantum potential from the modif\/ied
Hamilton--Jacobi equation and preserve its standard form as in
classical mechanics. The Wigner representation in phase space
which is obtained by a canonical transformation in EPS is found to
be the appropriate representation in which the quantum potential
is removed for these potentials. Generalization of the technique
for arbitrary potentials is being under consideration.

\appendix

\section{Appendix}

For a given Hamiltonian $H(p,q)$ one may consider
equation~\eqref{eq2}
 as a dif\/ferential equation for ${\cal L}^q$. For
harmonic potential,
\[
H(p,q)=\frac {1}{2m} p^2+\frac{1}{2}kq^2,\qquad k=m\omega^2,
\]
one obtains
\[
{\cal L}^q(q, \dot q)=\frac{1}{2}m\dot q^2-\frac{1}{2}kq^2,
\]
as the solution. In the same manner, using equation~\eqref{eq3},
one obtains
\[
{\cal L}^p(p, \dot p)=\frac{p^2}{2m}-\frac{1}{2k}\dot p^2,
\] as a possible solution.
Using Euler--Lagrange equation in the $q$ space, one gets,
\[
m\ddot{q}=-kq.
\] For an initial condition, say,
\[
q(0)=0, \qquad \dot q(0)=\dot q_0,
\]
the solution is
\[
q(t)=\frac{1}{\omega}\dot q_0\sin(\omega t) \qquad \mbox{and}
\qquad \dot q(t)=\dot q_0\cos(\omega t).
\]
For arbitrary initial conditions the solution in the $p$ space is
independent of the solution in the $q$ space. However, in order to
describe the same oscillator solution, the initial conditions
assumed for the oscillator in the $q$ space must be translated
into the $p$ space. This can be done using the relations that
instantaneously couples the $q$ space variables to those in the
$p$ space, i.e.,
\[
p=\frac{\partial {\cal L}^q}{\partial \dot q}=m\dot q,
\]
which is the def\/inition of the canonical momentum (here, the
same as the mechanical momentum) and
\[
q=\frac{\partial {\cal L}^p}{\partial \dot p}=-\frac{\dot p}{k}
\qquad \mbox{or} \qquad \dot p=-kq,
\]
which is Newton law of motion. Thus, one obtains
\[
p(0)=m\dot q_0 \qquad \mbox{and}\qquad \dot p(0)=0,
\]
as initial conditions and
\[
p=m\dot q_0\cos \omega t, \qquad \dot p=-m\omega \dot q_0\sin
\omega t,
\]
as the solutions in the $p$ space describing the same oscillator
in the $q$ space. The same way could be followed for linear
potential while the variable $\dot p$ is now a constant.

\subsection*{Acknowledgements}

The f\/inancial support of Research Council of Zanjan University,
Zanjan, Iran is appreciated. I~also would like to thank Mr.
B.~Farnudi for editing the manuscript.

\LastPageEnding

\end{document}